\documentclass[onecolumn,superscriptaddress,showpacs,pra]{revtex4}
\usepackage{epsfig}
\usepackage{epstopdf}
\usepackage{delarray}
\usepackage{amsmath, amssymb}
\usepackage{bm}
\usepackage{graphicx}
\usepackage{placeins}
\usepackage{color}

\begin{document}
\title{Predicting the Ocean Currents using Deep Learning}

\author{Cihan Bay\i nd\i r}
\email{cihanbayindir@gmail.com}
\affiliation{Associate Professor, Engineering Faculty, \.{I}stanbul Technical University, 34469 Maslak, \.{I}stanbul, Turkey. \\
						 Adjunct Professor, Engineering Faculty, Bo\u{g}azi\c{c}i University, 34342 Bebek, \.{I}stanbul, Turkey. \\
						 International Collaboration Board Member, CERN, CH-1211 Geneva 23, Switzerland.}

\begin{abstract}
In this paper, we analyze the predictability of the ocean currents using deep learning. More specifically, we apply the Long Short Term Memory (LSTM) deep learning network to a data set collected by the National Oceanic and Atmospheric Administration (NOAA) in Massachusetts Bay between November 2002-February 2003. We show that the current speed in two horizontal directions, namely u and v, can be predicted using the LSTM. We discuss the effect of training data set on the prediction error and on the spectral properties of predictions. Depending on the temporal or the spatial resolution of the data, the prediction times and distances can vary, and in some cases, they can be very beneficial for the prediction of the ocean current parameters. Our results can find many important applications including but are not limited to predicting the tidal energy variation, controlling the current induced vibrations of marine structures and estimation of the wave blocking point by the chaotic oceanic current and circulation.

\end{abstract}
\maketitle


\section{\label{sec:level1}Introduction}
Ocean currents and circulations are one of the very important phenomena observed in the ocean hydrodynamics and they have attracted a lot of attention throughout the historical development of ocean science \cite{Richardson, Peterson}. These currents and circulations can be generated by many different mechanisms. These mechanisms include but are not limited to winds \cite{Munk, Sverdrup47}, tides \cite{Parker}, waves \cite{Stommel}, changes in underwater topography \cite{Tomczak} and buoyancy fluxes \cite{Stommel, Wust, Sverdrup42, Siedler}. There are different ways of classifying the currents. According to their forcing mechanisms, they may be classified as wind driven or thermohaline. One other possible classification is according to the depth of occurrence, currents may be classified as the surface or the subsurface currents. Majority of the currents occurring on the ocean surface are wind driven, however thermohaline currents are driven by the heat and salt concentration differences. Sinking of denser saltier water at colder parts of the ocean drives subsurface thermohaline currents. Yet another possible classification is to classify the currents according to their occurrence locations, i.e. the currents flowing some distance away the shore can be classified as an offshore current and the ones closer to shore can be classified as the inshore current. From a time series analysis point of view, a more appropriate classification is to classify them as periodic or mean currents. The currents which have changing speed and directions cyclically at regular intervals can be classified as periodic currents. However mean circulations in the ocean have mean parts which experience no or relatively little changes in time and space.

Ocean currents and circulations have very important effects in the marine environment. They are one of the driving mechanisms of the mass and heat transfer on earth. They can heat up or cool down the oceans and can transport biological agents, nutrients and chemicals. They may lead to many engineering problems for the offshore engineering structures and marine travel, such as excessive vibrations, increase of transit times and accidents risks in strategic waterways, such as the Bosphorus \cite{Oguz, Jarosz}. Additionally, they may lead to formation of rogue waves due to their wave blocking effect \cite{Bayindir, BayPRE1, BayPRE2, BayPLA}.  They are the main driving mechanism of the tidal energy converters \cite{Magagna,Uihlein, Lynn}.

Due to the reasons, which some are summarized above, the prediction of ocean currents and circulation in the marine environment is a vital problem for the safety of the marine operations and structures. Various attempts are made for the prediction of the different ocean processes including the ocean currents and circulations. Various wave models and time series methods are utilized for the prediction of the ocean wave height and energy in \cite{Roulston, Reikard, Baysci}. The predictability of the near surface winds using Kalman filtering technique is discussed in \cite{Malmberg}. A transfer function based method was proposed in \cite{Ho} for the wave height forecast. The usage of artificial neural networks are discussed for the prediction of waves is studied in \cite{Londhe}. A genetic programming approach is used for the real time wave forecasting in \cite{Gaur}. More recently, the prediction of wave conditions using a machine learning approach is proposed in \cite{James_machinewave}. Again a machine learning approach is proposed in \cite{Jiang_ml_thermocline} for the prediction of thermocline parameters. A statistical machine learning approach is utilized in \cite{Hollinger} for the prediction of ocean processes. Spatio-temporal prediction of the ocean currents using Gaussian processes is discussed in \cite{Sarkar}. Another approach was to apply the deep learning to ocean data inference and subgrid parameterization \cite{Bolton}.

In this study, we discuss and analyze the predictability of the ocean currents using a deep learning approach \cite{MacKay}. To be more specific, we apply the LSTM deep learning network to a ocean current data set collected by NOAA in Massachusetts Bay between November 2002-February 2003. We discuss the performance of the LSTM for the prediction of the ocean current speed data and discuss the root-mean-square (rms) error and spectral properties of predictions. We also investigate the effect of the length of the training data set on predictions. We discuss our findings and comment on our results.

\section{\label{sec:level2}Methodology}

\subsection{Review of the Long Short Term Memory} 
\begin{figure}[htb!]
\begin{center}
   \includegraphics[width=3.7in]{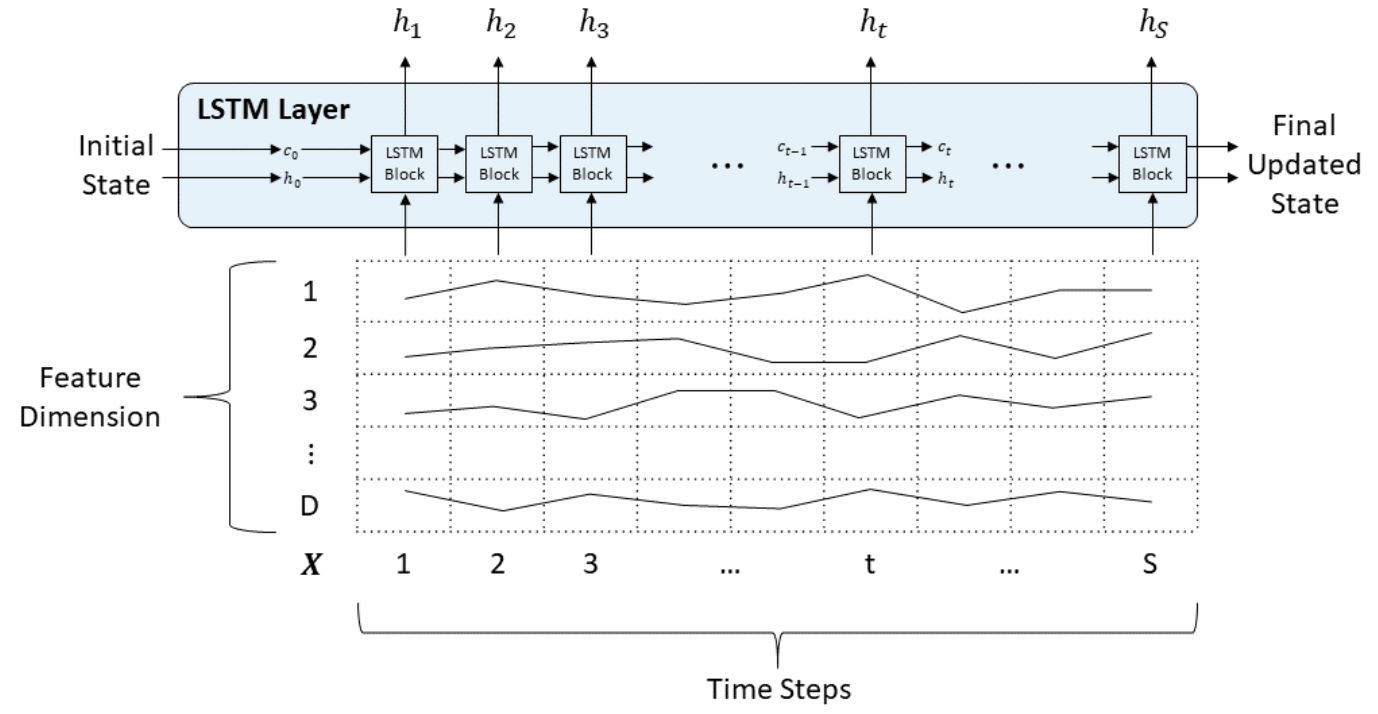}
  \end{center}
\caption{\small LSTM layer architecture \cite{Hochreiter}.}
  \label{figLSTM_layer}
\end{figure}

One of the most commonly used deep learning networks is the LSTM network, which was introduced in \cite{Hochreiter}. One of the possible uses of the LSTM is the prediction of the various time series data \cite{Hochreiter, Greff}. In order to predict the values of the data at the future time steps, a sequence-to-sequence regression LSTM network can be trained. For the LSTM network, training sequences with one time step shifted values are used as the responses \cite{Hochreiter}, thus the LSTM network learns to predict the values of the next time step at every time step of the training sequence \cite{Hochreiter}. The LSTM layer architecture is shown in Fig.~\ref{figLSTM_layer}.

\begin{figure}[htb!]
\begin{center}
   \includegraphics[width=3.7in]{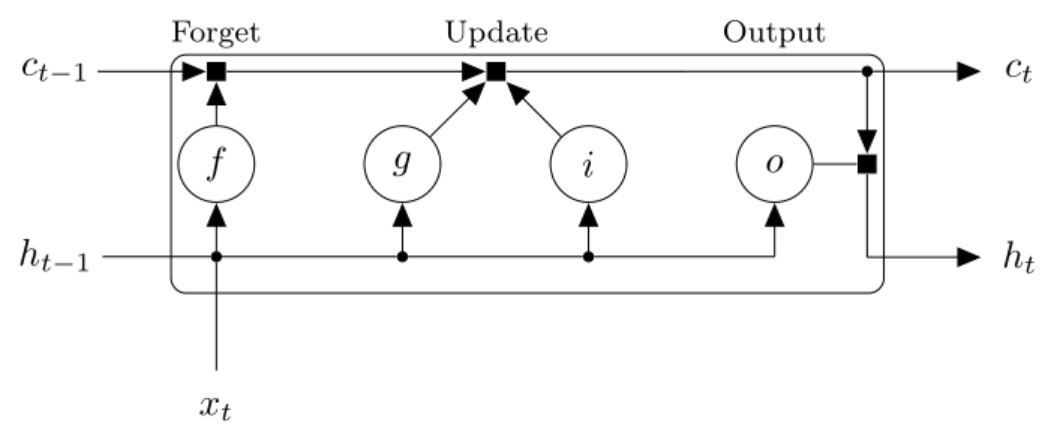}
  \end{center}
\caption{\small LSTM gates \cite{Hochreiter}.}
  \label{figLSTM_gates}
\end{figure}

In the LSTM layer architecture, the flow of a time series having D features through an LSTM layer is illustrated. In this architecture, h shows the output and c shows the cell state \cite{Hochreiter}.
The learned information at the previous time steps is contained in the cell state. The information is either added or removed from the cell state at each time step of the input sequence. The gates shown in Fig.~\ref{figLSTM_gates}, controls these updates.

\begin{figure}[htb!]
\begin{center}
   \includegraphics[width=2.2in]{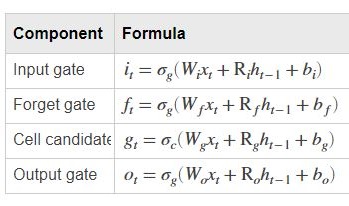}
  \end{center}
\caption{\small LSTM components and formula \cite{Hochreiter}.}
  \label{figLSTM_components}
\end{figure}

The letters i, f, g, o in this figure denote the input, forget gates, cell candidate and the output gate, respectively  \cite{Hochreiter}. At time step t, the cell state is computed by
\begin{equation}
c_t = f_t \otimes c_{t-1} + i_t \otimes g_{t}
\label{eq01}
\end{equation}
In here, $\otimes$ is the Hadamard product (element wise multiplication)  \cite{Hochreiter}. At time t, the hidden state is computed using
\begin{equation}
h_t = o_t \otimes \sigma_c(c_t)
\label{eq02}
\end{equation}
where $\sigma_c$ is the state activation function. Among different possibilities, the tanh function is used for the state activation function throughout this study. The functions are shown in Fig.~\ref{figLSTM_components}
In here W shows the input weight, R shows the recurrent weight and b shows the bias \cite{Hochreiter}. $\sigma_g$ shows the gate activation function. A sigmoid function, $\sigma(x)=(1+e^{-x})^{-1}$ is used as the gate activation function \cite{Hochreiter}. In this approach, the training data is standardized to have zero mean and unit variance for a better fit. Then, the LSTM layer is specified to have 200 hidden units. The number of training epochs is selected as 250 and a gradient threshold value of 1 is used to avoid the divergence of the gradients. The predictions are started 
with an initial learning rate of 0.005, and after 200 epochs, it is dropped to 0.004. Then the predicted time series is unstandardized using the parameters discussed earlier. Additionally, in LSTM network approach one can also update the network state with observed values instead of predictions by utilizing the time steps between predictions which generally results in better prediction performance. All of the LSTM network based predictions in this work are performed using the parameters discussed above. Our paper focuses on discussing the usage of the deep learning for the prediction of ocean currents and circulations. Thus, the reader is referred to \cite{Hochreiter, Greff} for a more comprehensive discussion and the details of the LSTM network.

\section{Properties and Review of the Ocean Current Data}
\begin{figure}[htb!]
\begin{center}
   \includegraphics[width=4.7in]{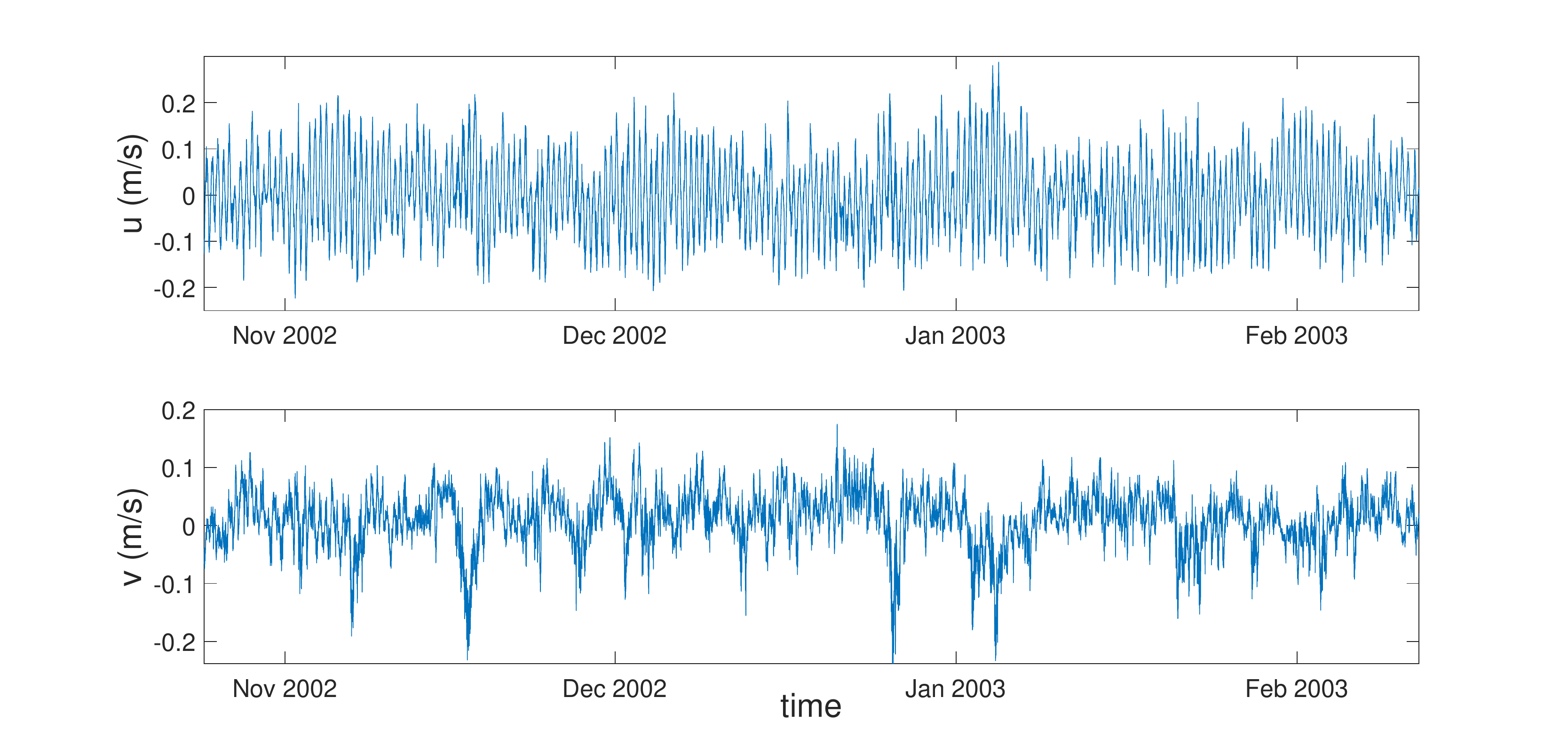}
  \end{center}
\caption{\small Ocean current speed time series  a) u (m/s) b) v (m/s). }
  \label{fig1}
\end{figure}

In this study, we use an experimental data set to test the performance of the LSTM deep learning network for the ocean current and circulation predictions. The data set is recorded by NOAA in North Atlantic in the Massachusetts Bay at the location $42.3773N, 70.7819W$. The experimental data was recorded using a VMCM type moored current meter between the dates $2002/10/24$ and $2003/02/12$. The seafloor depth at this specific location was $33.7m$ and the current meter was deployed at a depth of $23.5m$. The horizontal current velocities, namely u and v are measured at every 3 minutes 44 seconds during this experiment. The original data set and further information can be seen at NOAA's website \hyperref{https://www.nodc.noaa.gov/gocd/data/a0060062/info/gocd{\_}a0060062{\_}6964vm-a.html}. The time series of the 2D horizontal current velocities, u and v are depicted in Fig.~\ref{fig1}.

\section{\label{sec:level3}Results and Discussion}
 
In this section, we assess the performance of the LSTM deep learning network for the prediction of the ocean current speed data depicted in Fig.~\ref{fig1} in the preceding section. 

\begin{figure}[htb!]
\begin{center}
   \includegraphics[width=4.3in]{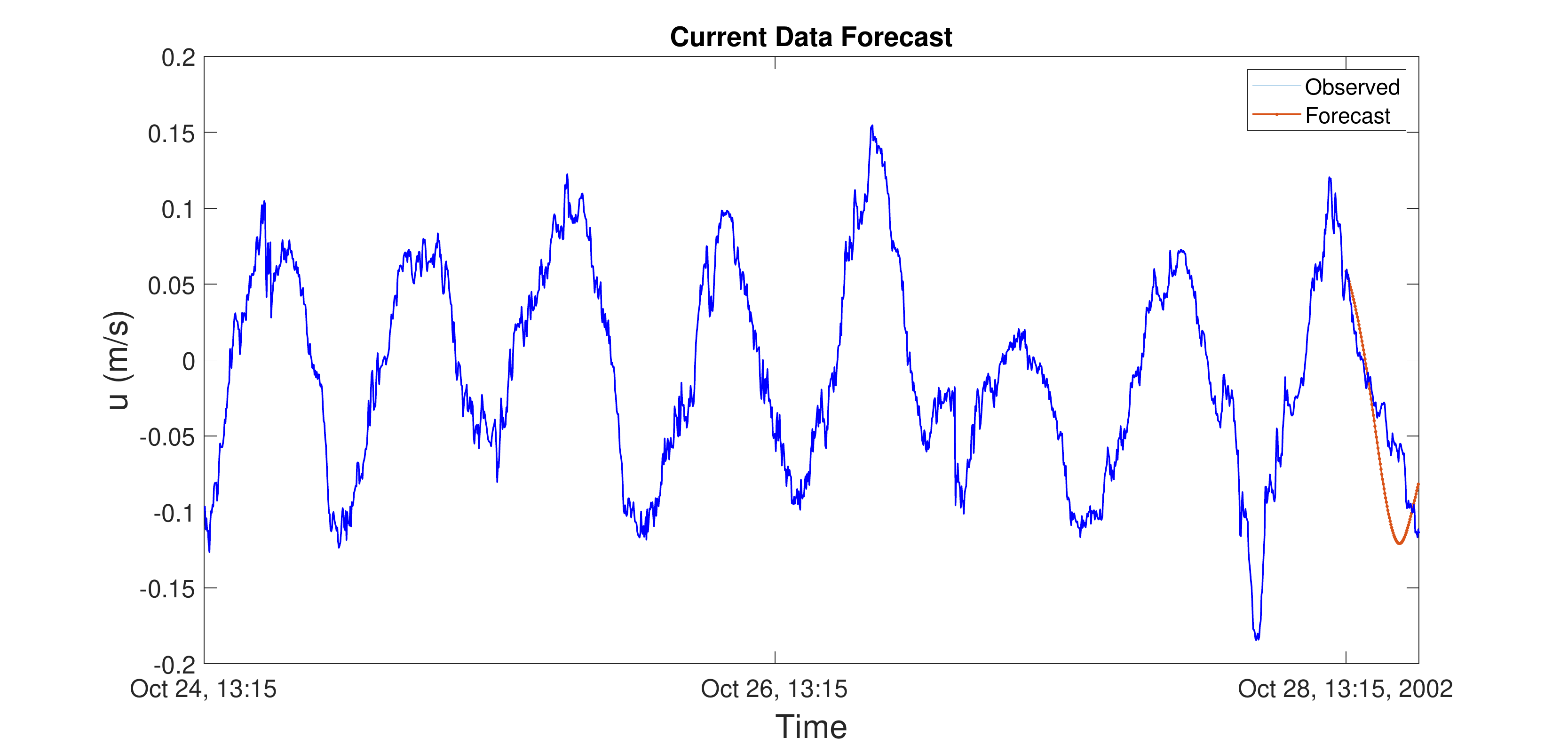}
  \end{center}
\caption{\small Observed and predicted time series of the first component of the current velocity (u) obtained using the initial 95 \% as the training sequence.}
  \label{fig2}
\end{figure}

\begin{figure}[htb!]
\begin{center}
   \includegraphics[width=4.3in]{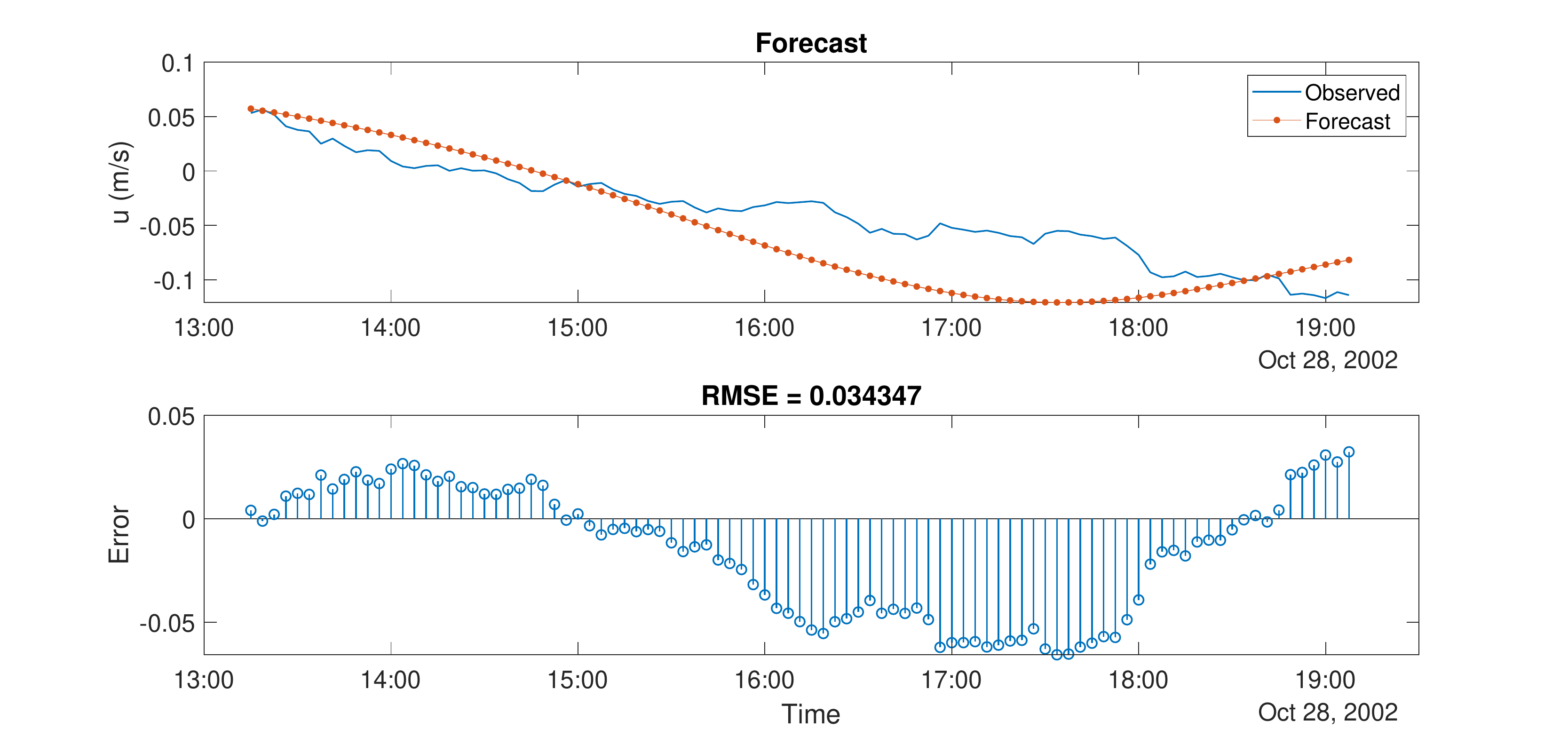}
  \end{center}
\caption{\small a) Comparisons between the observed and predicted time series of the first component of the current velocity (u) obtained using the initial 95 \% as the training sequence, b) the rms error of predictions.}
  \label{fig3}
\end{figure}

In Fig.~\ref{fig2}, the initial part of the first component of the original current speed data (u) between the dates October 24-October 28, 2002 is shown. In this prediction, we use the initial 95 \% of this part of the current speed (u) data set as the training sequence. The training data set (in blue) and the predicted time series (in red) are depicted in Fig.~\ref{fig2}. We compare the predicted and observed time series and depict the root-mean-square (rms) error in Fig.~\ref{fig3}. As one can realize from these figures, the LSTM deep learning network performs quite well for the prediction of the first component of ocean current speed, u. The predictions are better for the first few time steps. As depicted in Fig.~\ref{fig3}, the absolute value of the rms error remains less than $0.05m/s$ for the data having an absolute peak value around $0.15m/s$. It is useful to note that, the sampling interval of this data set is 3 minutes 44 seconds. An early prediction time on the order of few time steps can be very beneficial for many practices in marine science such as for tidal energy harvesting and vibration control, just to name a few. Depending on the temporal resolution of the data and with the advance of algorithms, this prediction times can vary and may be substantially improved.

\begin{figure}[htb!]
\begin{center}
   \includegraphics[width=4.3in]{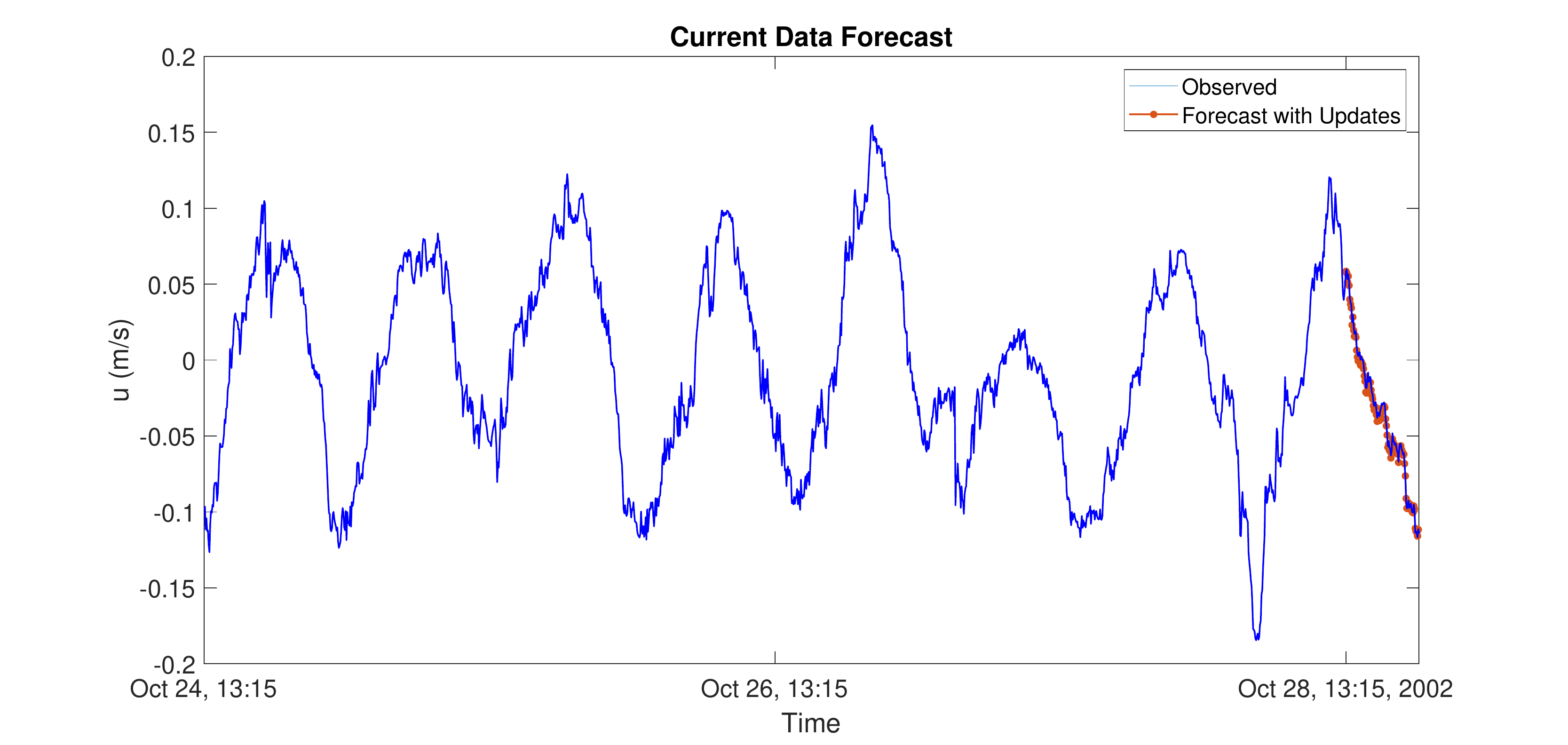}
  \end{center}
\caption{\small Observed and predicted time series of the first component of the current velocity (u) obtained using the initial 95 \% as the training sequence and using the updates.}
  \label{fig4}
\end{figure}

\begin{figure}[htb!]
\begin{center}
   \includegraphics[width=4.3in]{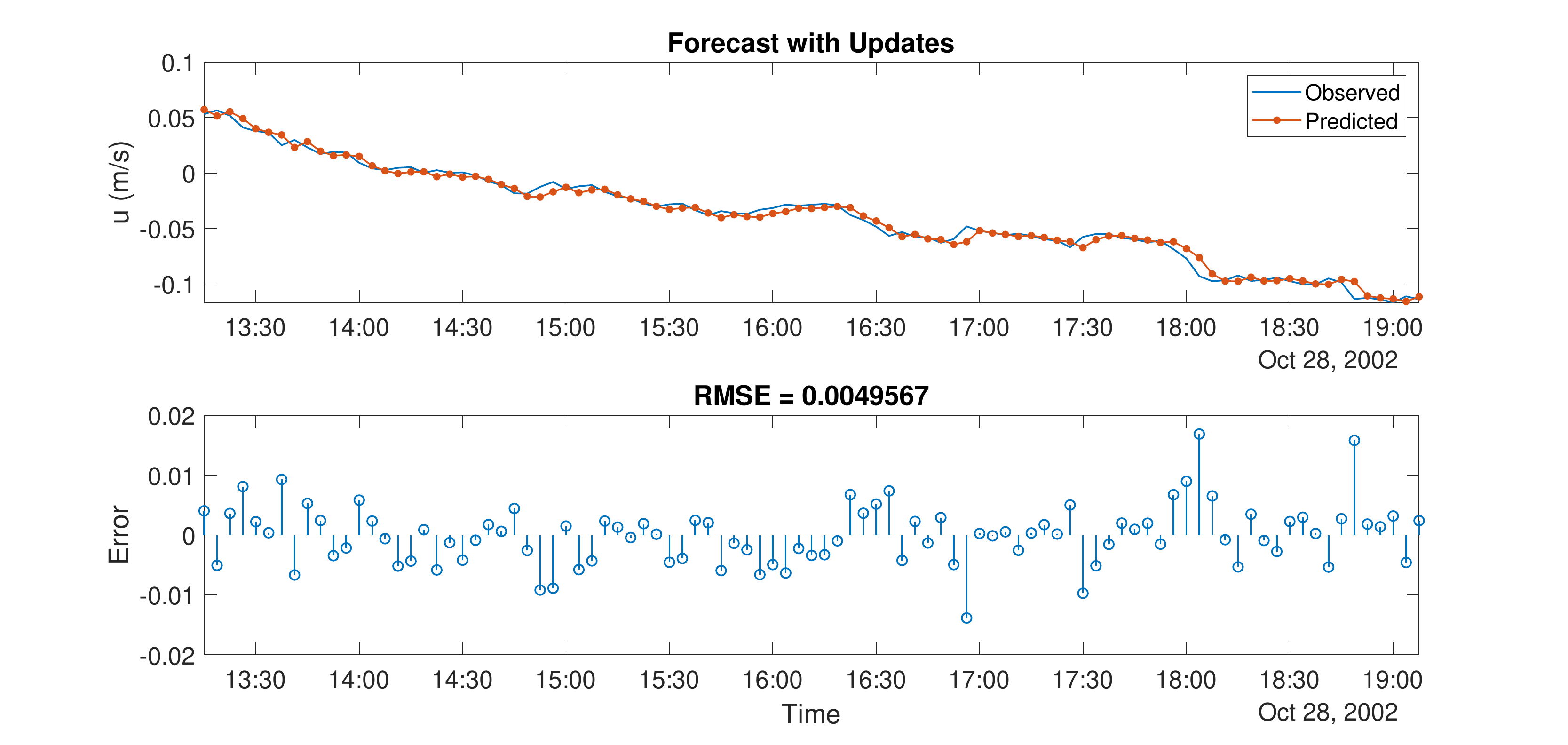}
  \end{center}
\caption{\small a) Comparisons between the observed and predicted time series of the first component of the current velocity (u) obtained using the initial 95 \% as the training sequence and using the updates, b) the rms error of predictions.}
  \label{fig5}
\end{figure}

Next, we turn our attention to discuss the effects of using the observed values instead of predicted ones which are used for updating the LSTM network. We plot the first component of the predicted ocean current speed (u) time series and the rms error for this case in Fig.~\ref{fig4} and Fig.~\ref{fig5}. The same data set depicted Fig.~\ref{fig2} is used and its initial 95 \% is used as the training sequence, as before. Checking Fig.~\ref{fig4} and Fig.~\ref{fig5}, it is possible to conclude that the LSTM prediction with updates leads to better predictions and rms error significantly reduces. For the same data set, the peak absolute value of the rms error becomes less than $0.02m/s$ for this case.

\begin{figure}[htb!]
\begin{center}
   \includegraphics[width=4.3in]{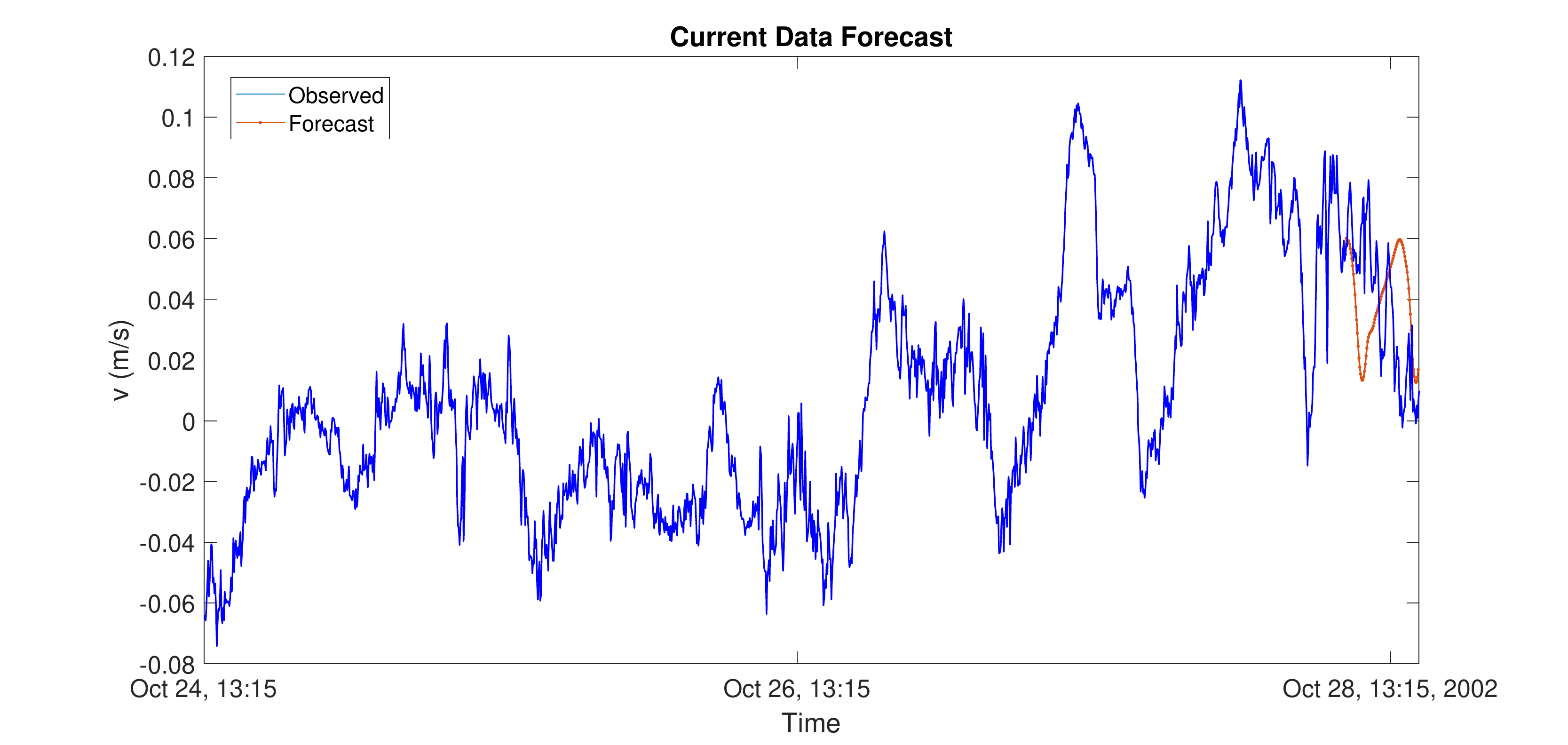}
  \end{center}
\caption{\small Observed and predicted time series of the second component of the current velocity (v) obtained using the initial 95 \% as the training sequence.}
  \label{fig6}
\end{figure}

\begin{figure}[htb!]
\begin{center}
   \includegraphics[width=4.3in]{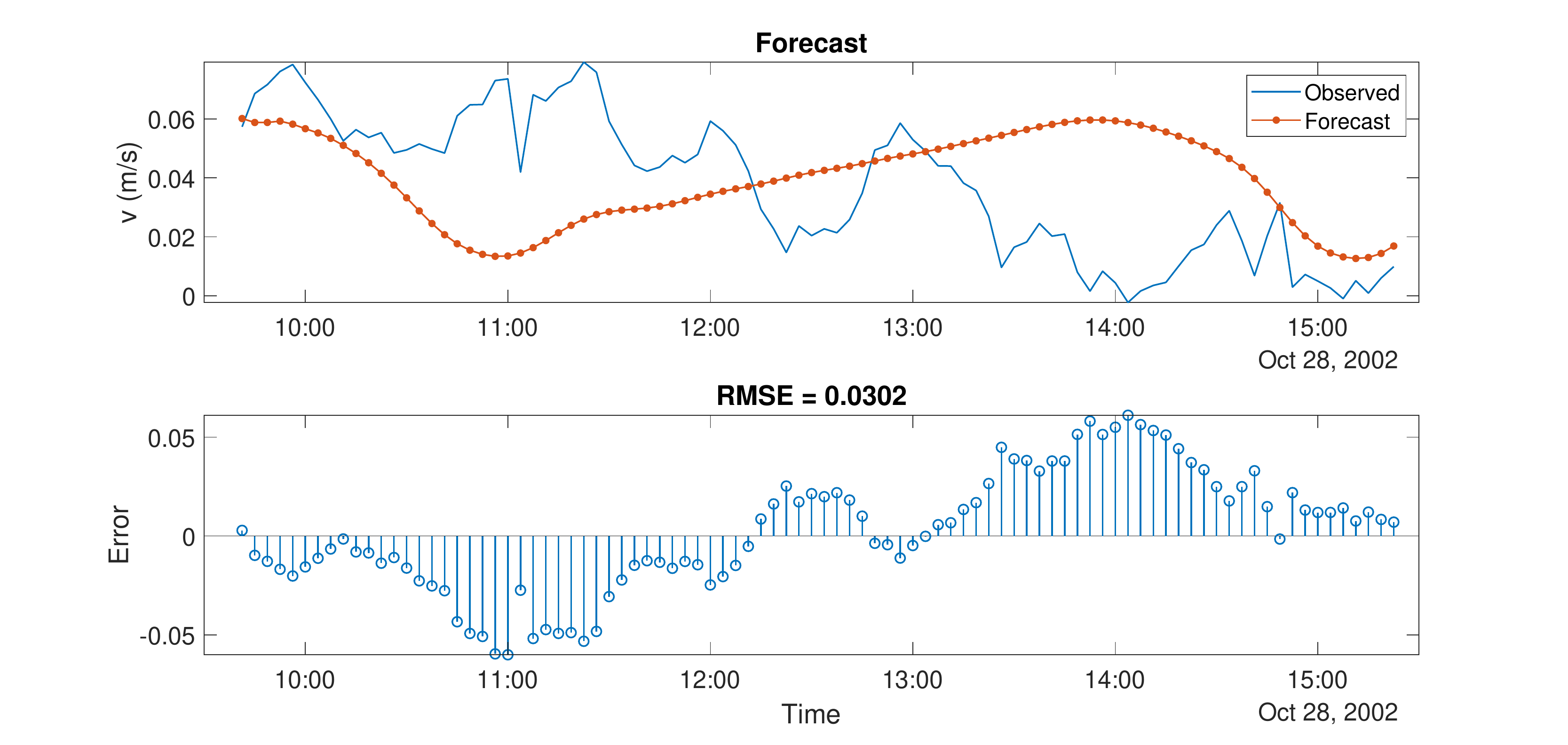}
  \end{center}
\caption{\small a) Comparisons between the observed and predicted time series of the second component of the current velocity (v) obtained using the initial 95 \% as the training sequence, b) the rms error of predictions.}
  \label{fig7}
\end{figure}

Focusing on the second component of the horizontal current velocity, namely v, we plot the predicted v time series and the rms error LSTM in Fig.~\ref{fig6} and Fig.~\ref{fig7}. The training sequence is selected as the initial 95 \%  of this part of the v time series. Checking these figures it is possible to realize that LSTM performs worse compared to the predictions of u. This is due to the difference in the behavior of time series. While u has a periodic steady state behavior, v exhibits a tendency of increase and does not have a steady state behavior. This structure of the data set makes the LSTM network predictions poorer. The absolute value of the rms error is on the order of $0.05m/s$ for the data having an absolute peak value around $0.10m/s$.

\begin{figure}[htb!]
\begin{center}
   \includegraphics[width=4.3in]{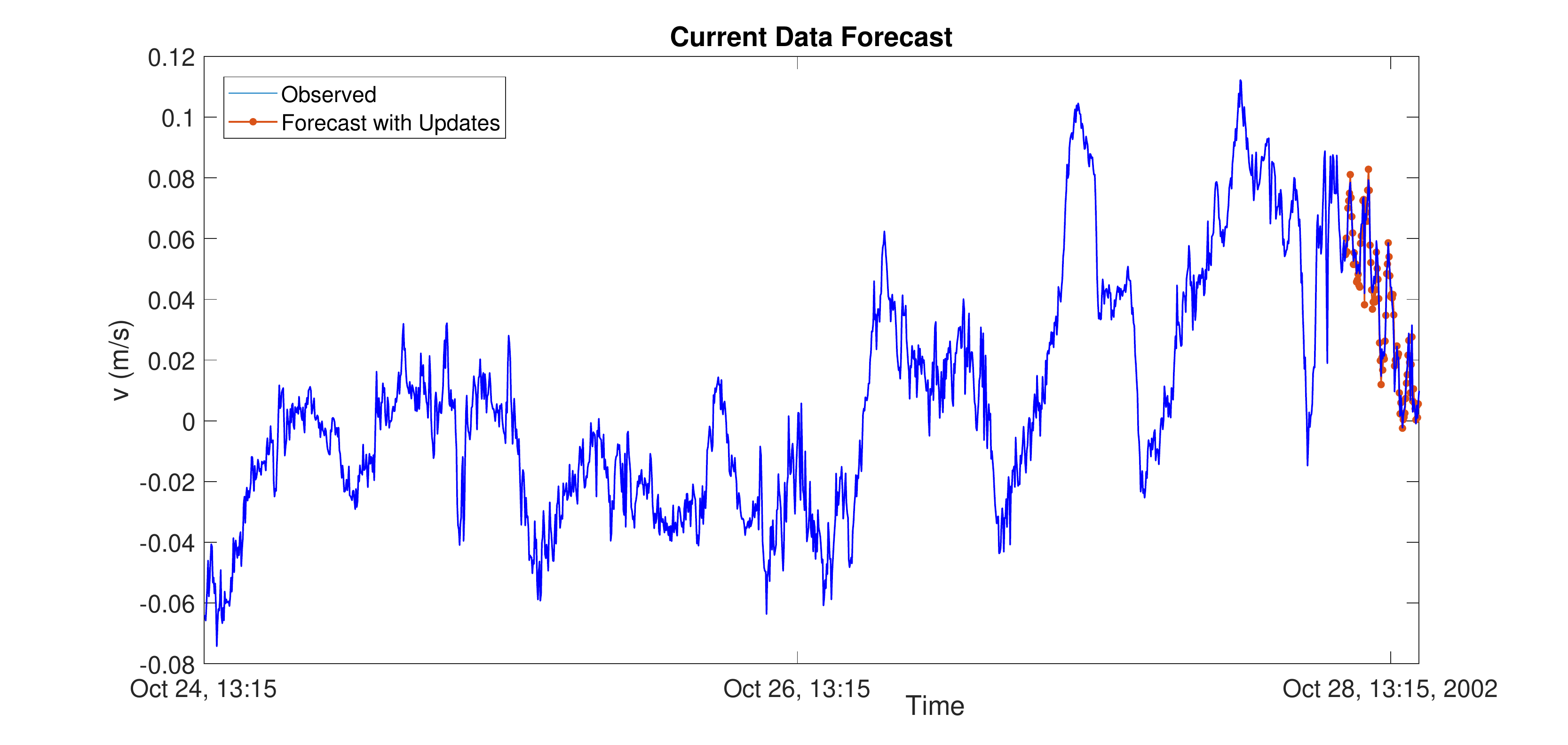}
  \end{center}
\caption{\small Observed and predicted time series of the second component of the current velocity (v) obtained using the initial 95 \% as the training sequence and using the updates.}
  \label{fig8}
\end{figure}

\begin{figure}[htb!]
\begin{center}
   \includegraphics[width=4.3in]{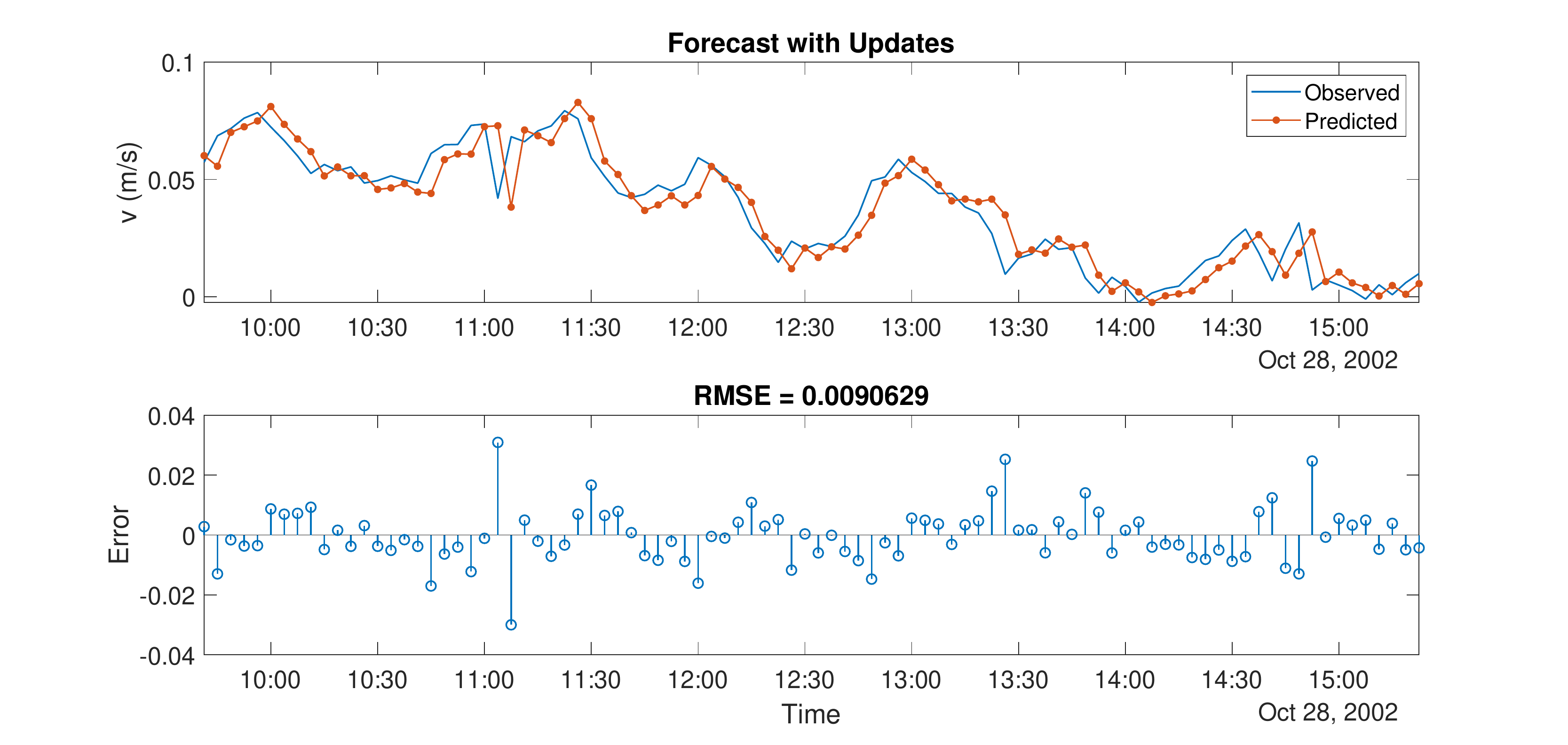}
  \end{center}
\caption{\small a) Comparisons between the observed and predicted time series of the second component of the current velocity (v) obtained using the initial 95 \% as the training sequence and using the updates, b) the rms error of predictions.}
  \label{fig9}
\end{figure}

In order to discuss the effects of updating the LSTM network with observed values, we repeat the same simulations for the case with updates and depict the related results in Fig.~\ref{fig8} and Fig.~\ref{fig9}. Comparing Fig.~\ref{fig6} and Fig.~\ref{fig7} with Fig.~\ref{fig8} and Fig.~\ref{fig9} respectively, it is possible to argue that using the observed values to update the LSTM network significantly improves the prediction performance. The peak absolute value of the rms error becomes less than $0.04m/s$ for this case. It is also useful to note that using the observed values as updates limits the prediction time to one time step. The value of time step is 3 min 44 seconds for the data set used, however, depending on the size of the sampling interval the prediction time can vary and can still be very beneficial. Additionally, it is possible to upsample or downsample the data and adaptive time stepping can be utilized to enhance the prediction time scales.

\begin{figure}[htb!]
\begin{center}
   \includegraphics[width=4.3in]{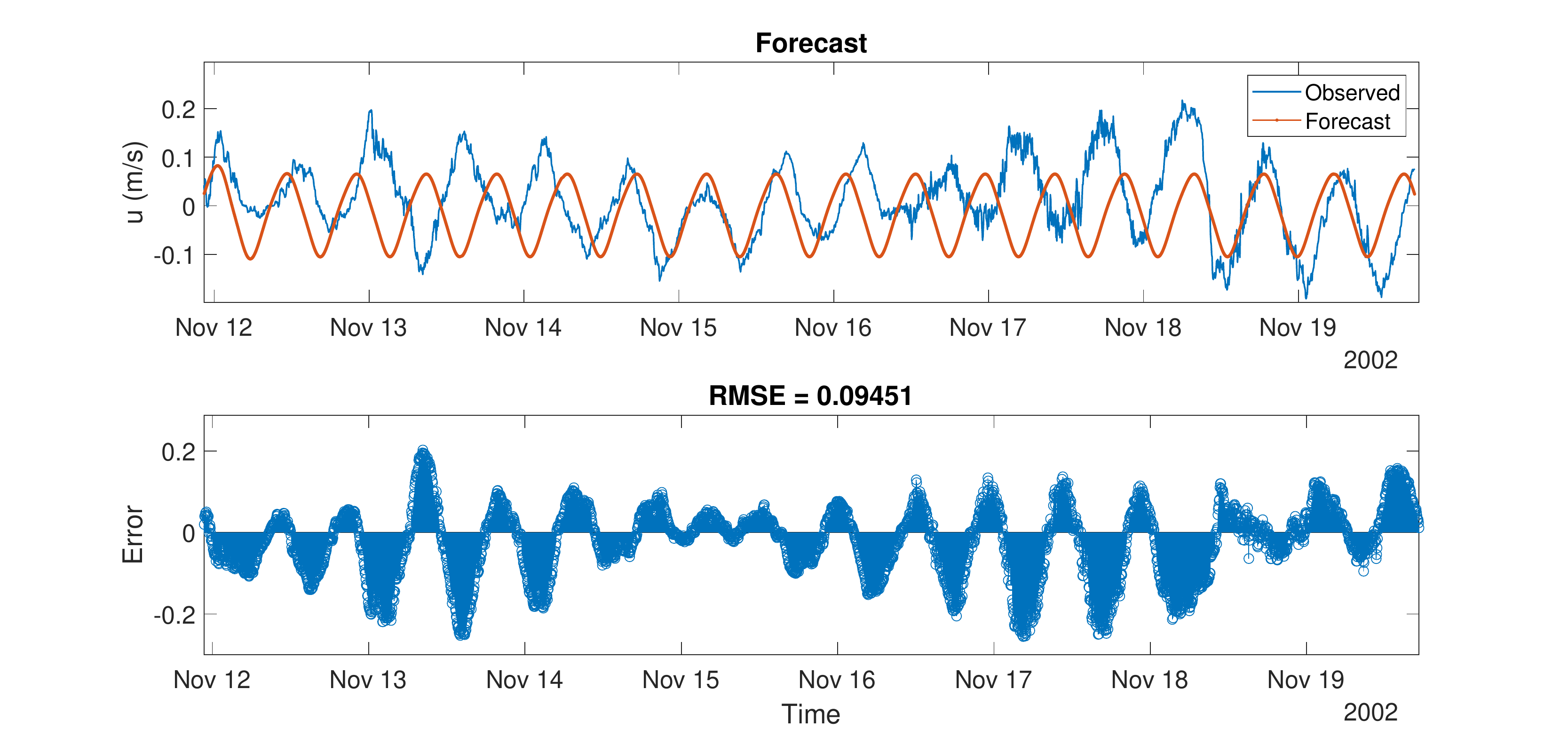}
  \end{center}
\caption{\small a) Comparisons between the observed and predicted time series of the first component of the current velocity (u) obtained using the initial 70 \% as the training sequence, b) the rms error of predictions.}
  \label{fig10}
\end{figure}

\begin{figure}[htb!]
\begin{center}
   \includegraphics[width=4.3in]{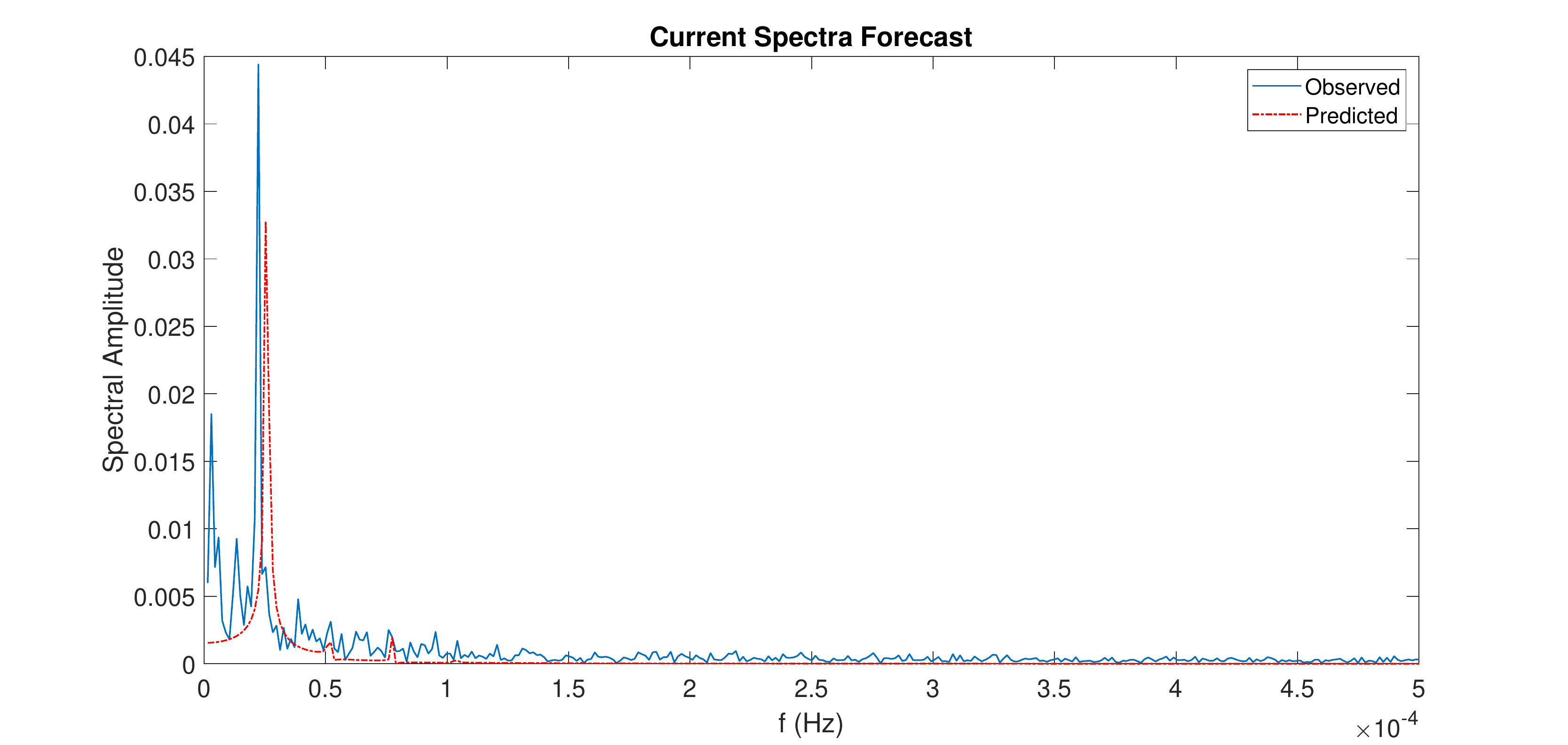}
  \end{center}
\caption{\small Comparison of the Fourier spectra of the observed and predicted time series of the first component of the current velocity (u) obtained using the initial 70 \% as the training sequence.}
  \label{fig11}
\end{figure}

\begin{figure}[htb!]
\begin{center}
   \includegraphics[width=4.3in]{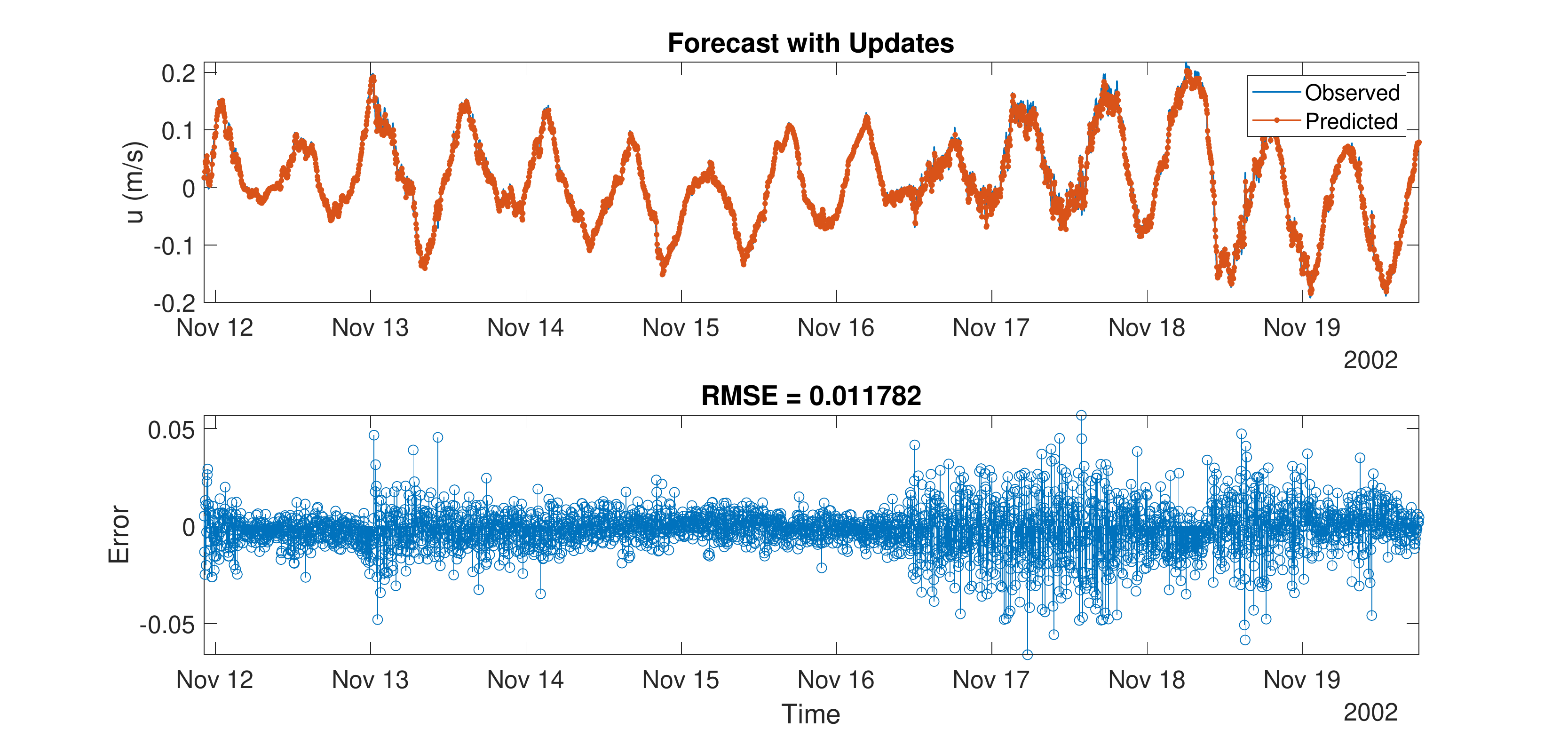}
  \end{center}
\caption{\small a) Comparisons between the observed and predicted time series of the first component of the current velocity (u) obtained using the initial 70 \% as the training sequence and using the updates, b) the rms error of predictions.}
  \label{fig12}
\end{figure}

\begin{figure}[htb!]
\begin{center}
   \includegraphics[width=4.3in]{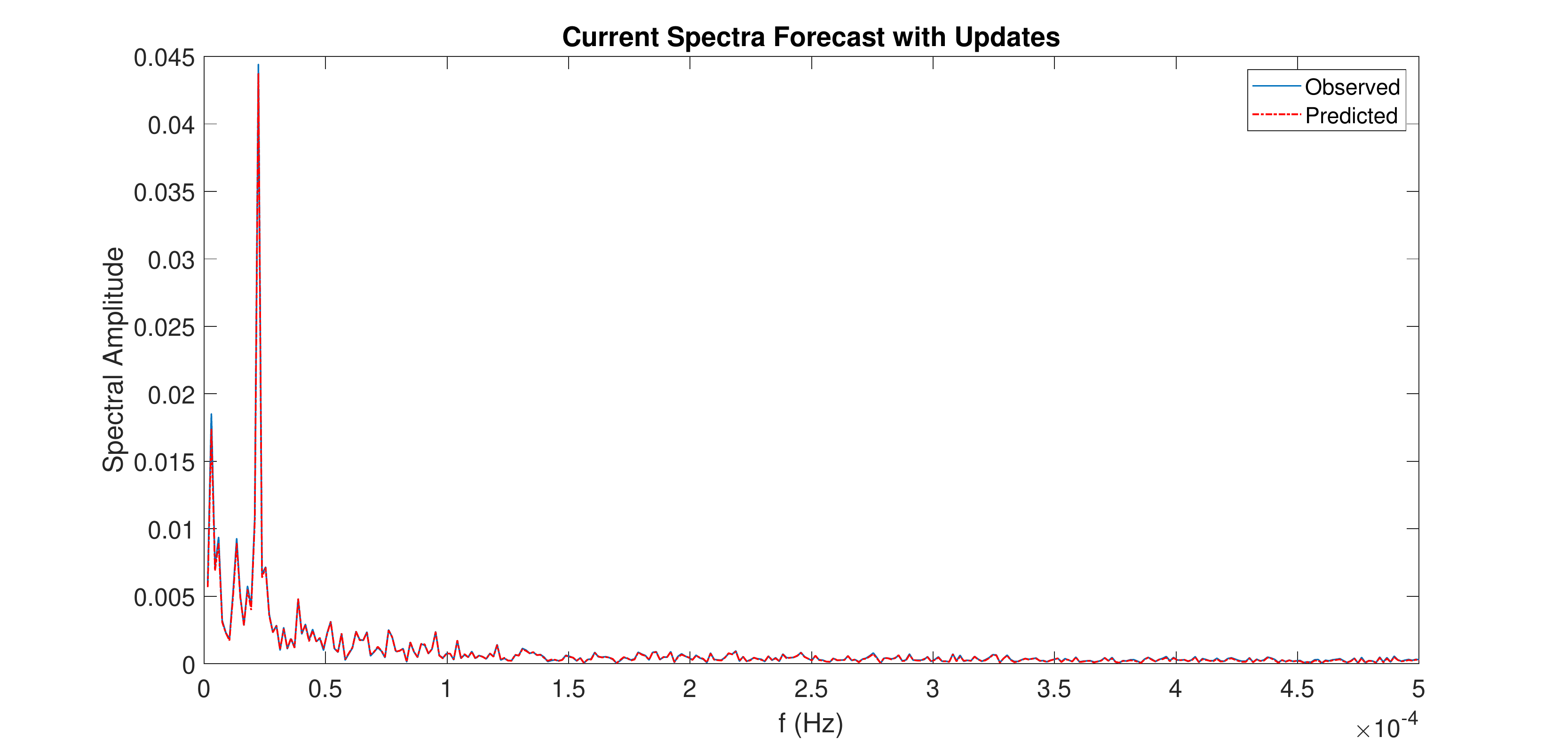}
  \end{center}
\caption{\small Comparison of the Fourier spectra of the observed and predicted time series of the first component of the current velocity (u) obtained using the initial 70 \% as the training sequence and using the updates.}
  \label{fig13}
\end{figure}

Next, we investigate the effects of the size of the LSTM training data set and the spectral properties of predictions obtained by the LSTM network. With this motivation, we use the initial part of time series between the dates October 24-November 19, 2002. We select the initial \% 70 of this part of the time series data as the training data set and predict the remaining part using the LSTM network. Results for the first component of the current speed data (u) with no updates are depicted in Fig.~\ref{fig10} and its spectrum is depicted in Fig.~\ref{fig11}. The results for u with updates are depicted in Fig.~\ref{fig12} and its spectrum is depicted in Fig.~\ref{fig13}.

Checking Figs.~\ref{fig10}-\ref{fig13}, it is possible to argue that LSTM with no updates produces quite sufficient results for the prediction of the ocean current speed. When the observed values are used as updates, the results significantly improve and predictions becomes excellent for many marine operations and marine engineering purposes. The depicted Fourier spectra in Fig.~\ref{fig12} and Fig.~\ref{fig13} clearly show that the predicted time series with LSTM network has a higher peak frequency and its bandwidth is significantly limited compared to the observations, when no observations are used as updates. Again, when the observations are used as updates, the Fourier spectrum of the predicted time series by the LSTM network is in excellent agreement with the spectrum of the observations. Comparing Fig.~\ref{fig1} and Fig.~\ref{fig10}, it is also possible to state that when the training sequence is longer the prediction performance of the LSTM network significantly improves, as expected.

\begin{figure}[htb!]
\begin{center}
   \includegraphics[width=4.3in]{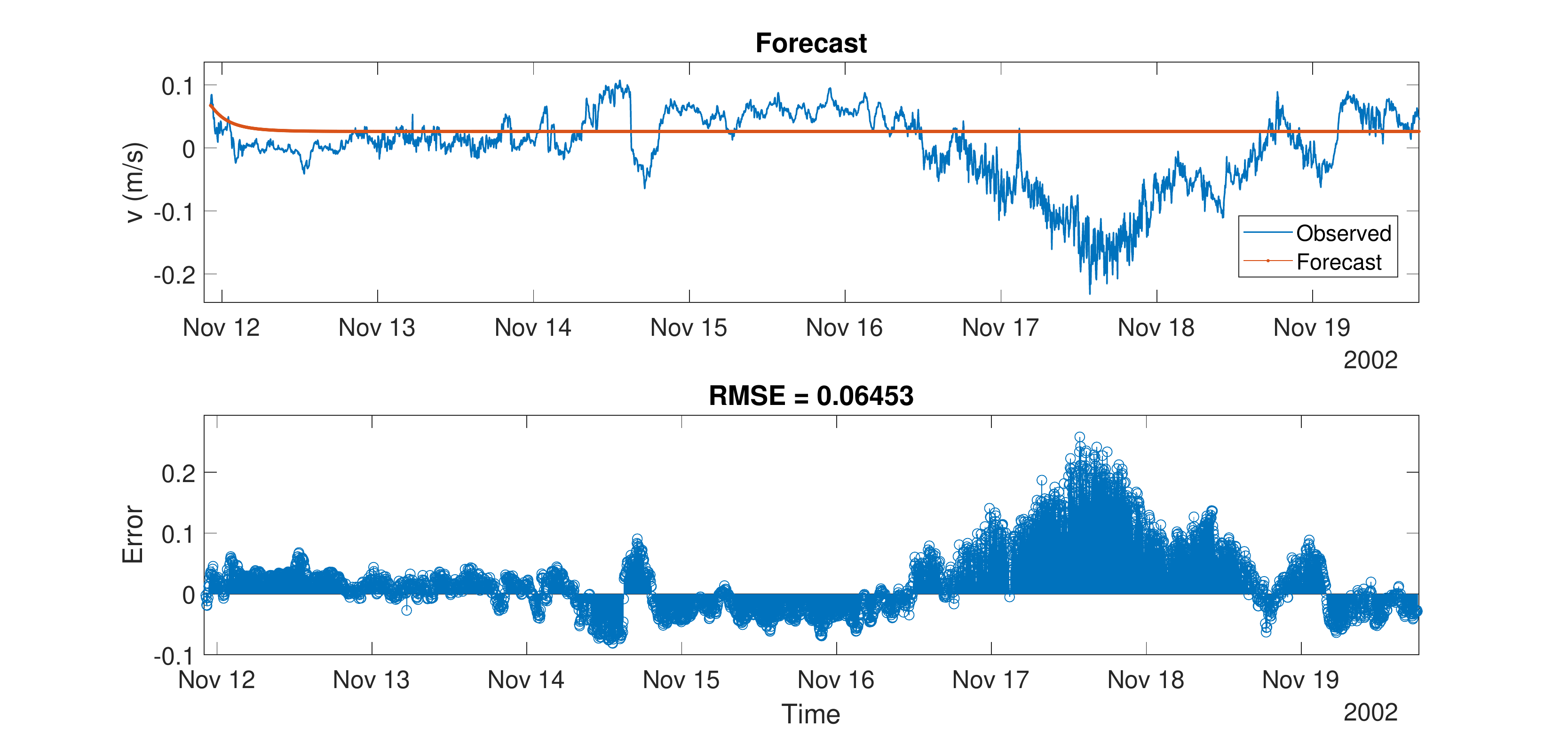}
  \end{center}
\caption{\small a) Comparisons between the observed and predicted time series of the second component of the current velocity (v) obtained using the initial 70 \% as the training sequence, b) the rms error of predictions.}
  \label{fig14}
\end{figure}

\begin{figure}[htb!]
\begin{center}
   \includegraphics[width=4.3in]{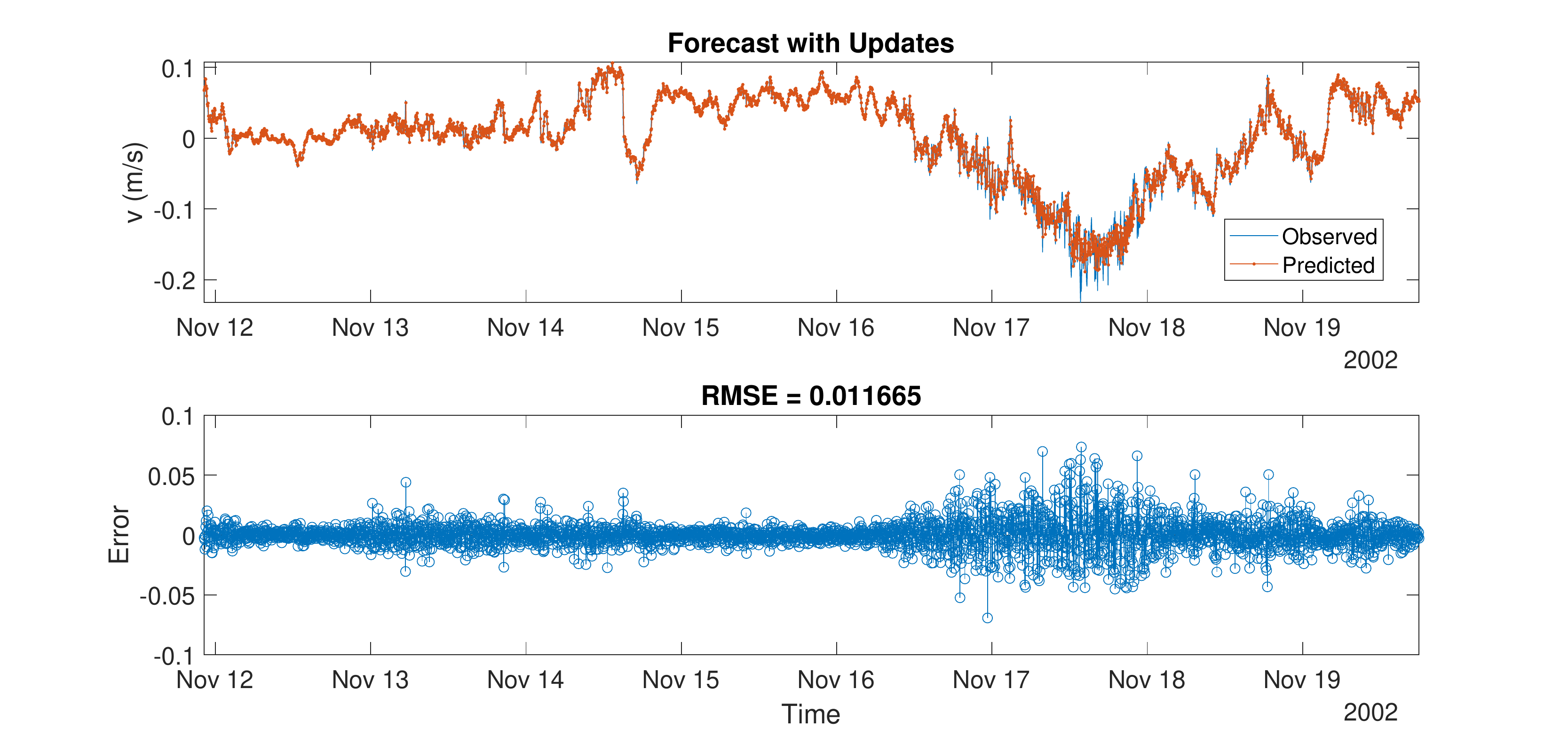}
  \end{center}
\caption{\small a) Comparisons between the observed and predicted time series of the second component of the current velocity (v) obtained using the initial 70 \% as the training sequence and using the updates, b) the rms error of predictions.}
  \label{fig15}
\end{figure}

\begin{figure}[htb!]
\begin{center}
   \includegraphics[width=4.3in]{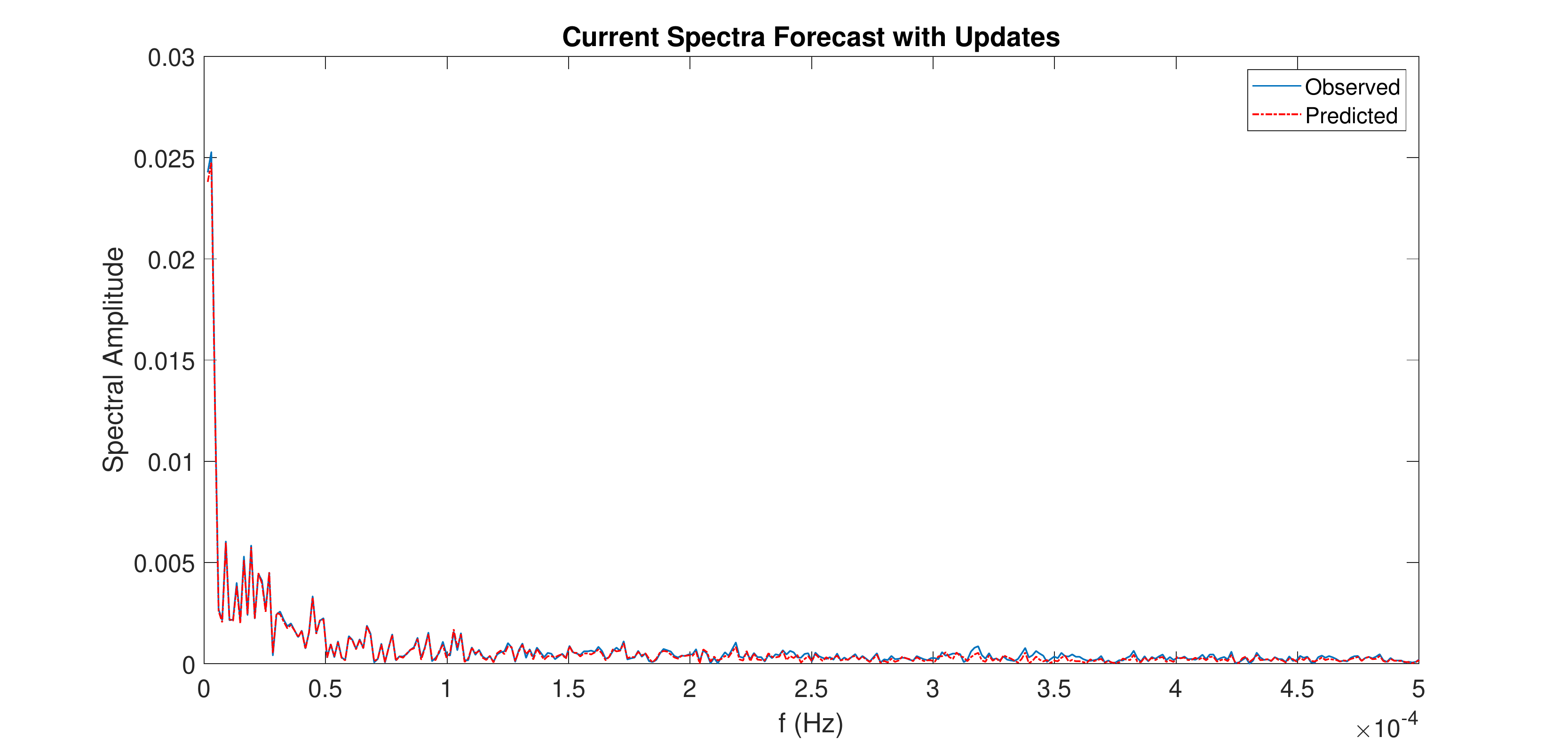}
  \end{center}
\caption{\small Comparison of the Fourier spectra of the observed and predicted time series of the second component of the current velocity (v) obtained using the initial 70 \% as the training sequence and using the updates.}
  \label{fig16}
\end{figure}

Lastly, we turn our attention again to the to the second component of the current speed time series, namely v. Again, we use the initial part of v time series between the dates October 24-November 19, 2002 and as before we select the initial \% 70 of this part of the data set as our training sequence. The predicted v time series with no updates is depicted in Fig.~\ref{fig14} and the predicted v time series with updates is depicted in Fig.~\ref{fig15}. Checking these figures, it is possible to argue that the prediction with no updates is poor and can not represent the data after few time steps. This mismatch is a result of using a shorter training sequence as well as the unsteady nature of the v time series. However, when the observed values are used as updates in the LSTM network, the predicted time series and its spectrum become excellent for many possible engineering uses as depicted in Figs.~\ref{fig15} and \ref{fig16}.

\section{\label{sec:level1}Conclusion and Future Work}
In this paper, we have discussed the predictability of the ocean current speed time series by deep learning. Specifically, we have investigated the applicability of the LSTM deep learning network to the ocean current speed time series and discussed its prediction performance. The experimental data set used in our study was collected by NOAA in Massachusetts Bay between the dates November 2002-February 2003. Implementing the LSTM network based deep learning approach on this data set, we have analyzed the rms error and the Fourier spectra of the predicted time series. We have showed that, depending on the steady state behavior of the predicted time series, the LSTM network with no updates can be very successful in the prediction of the ocean current speed data, and may allow for accurate predictions at least a few time steps in advance. We showed that the LSTM network predictions with no updates have a higher peak frequency compared to the observed spectra. However, when the observed values are used as updates in the LSTM network, the prediction accuracy significantly develops. We have also showed that longer training sequences result in better prediction accuracy of the ocean current time series. Depending on the sampling interval of the data, the prediction times can significantly vary. Thus, LSTM network based deep learning approach proposed in this paper can be used to predict the ocean current time series data with very beneficial prediction times. Our results can be generalized for the prediction of other oceanic and atmospheric time series data including but are not limited to Gulf Stream, North Equatorial, Agulhas currents and East Wind drift. Possible usage areas include but are not limited to predicting the tidal energy variation, controlling the current induced vibrations of marine structures and estimation of the wave blocking point by the chaotic oceanic current and circulation.

\end{document}